\documentclass[prb,aps,twocolumn]{revtex4-1}
\usepackage{amsmath} 
\usepackage{amsthm} 
\usepackage{amssymb}	
\usepackage{graphicx} 
\usepackage[usenames]{color}
\usepackage{array} 
\usepackage{paralist} 
\usepackage{amsfonts}
\usepackage{mathtools}
\usepackage{times}
\usepackage{braket}
\usepackage[colorlinks=true,linkcolor=blue,citecolor=blue,breaklinks]{hyperref}
\normalfont

\renewcommand{\vec}[1]{\ensuremath{\mathbf{#1}}} 
 
\newcommand{\avg}[1]{\langle #1 \rangle} 
 
\let\baraccent=\= 
\renewcommand{\=}[1]{\stackrel{#1}{=}} 

\theoremstyle{definition}

\theoremstyle{remark}

\begin{document}
\newcommand\bbone{\ensuremath{\mathbbm{1}}}
\newcommand{\ul}{\underline}
\newcommand{\vl}{v_{_L}}
\newcommand{\vc}{\mathbf}
\newcommand{\be}{\begin{equation}}
\newcommand{\ee}{\end{equation}}
\newcommand{\bk}{{{\bf{k}}}}
\newcommand{\bK}{{{\bf{K}}}}
\newcommand{\cE}{{{\cal E}}}
\newcommand{\bQ}{{{\bf{Q}}}}
\newcommand{\br}{{{\bf{r}}}}
\newcommand{\bg}{{{\bf{g}}}}
\newcommand{\bG}{{{\bf{G}}}}
\newcommand{\hbr}{{\hat{\bf{r}}}}
\newcommand{\bR}{{{\bf{R}}}}
\newcommand{\bq}{{\bf{q}}}
\newcommand{\hx}{{\hat{x}}}
\newcommand{\hy}{{\hat{y}}}
\newcommand{\hd}{{\hat{\delta}}}
\newcommand{\bea}{\begin{eqnarray}}
\newcommand{\eea}{\end{eqnarray}}
\newcommand{\beal}{\begin{align}}
\newcommand{\eeal}{\end{align}}
\newcommand{\ra}{\rangle}
\newcommand{\la}{\langle}
\renewcommand{\tt}{{\tilde{t}}}
\newcommand{\upa}{\uparrow}
\newcommand{\dna}{\downarrow}
\newcommand{\bS}{{\bf S}}
\newcommand{\vS}{\vec{S}}
\newcommand{\dg}{{\dagger}}
\newcommand{\pdg}{{\phantom\dagger}}
\newcommand{\tphi}{{\tilde\phi}}
\newcommand{\cf}{{\cal F}}
\newcommand{\ca}{{\cal A}}
\renewcommand{\ni}{\noindent}
\newcommand{\ct}{{\cal T}}
\newcommand{\zp}[1]{ { \color{red} \footnotesize ZP\; \textsf{\textsl{#1}} } }

\title{Mean field study of the topological Haldane-Hubbard model of spin-$1/2$ fermions}
\author{V. S. Arun,$^{1}$ R. Sohal,$^{1}$ C. Hickey,$^{1}$, and A. Paramekanti$^{1,2}$}
\affiliation{$^1$Department of Physics, University of Toronto, Toronto, Ontario M5S 1A7, Canada}
\affiliation{$^2$Canadian Institute for Advanced Research, Toronto, Ontario M5G 1Z8, Canada}
\begin{abstract}
Motivated by exploring the effect of interactions on Chern
insulators, and by recent experiments realizing topological bands for ultracold atoms in synthetic gauge fields, we study
the honeycomb lattice Haldane-Hubbard model of spin-$1/2$ fermions. Using an unrestricted
mean field approach, we map out the instability of the topological band
insulator towards magnetically ordered insulators which emerge with increasing Hubbard repulsion. 
In addition to the topological N\'eel phase, we recover various chiral noncoplanar magnetic 
orders  previously identified within a strong coupling approach. We compute the band structure
of these ordered phases, showing that the triple-$Q$ tetrahedral phase harbors topological Chern
bands with large Chern numbers.
\end{abstract}
\maketitle

\section{Introduction}

Theory and experiments on spin-orbit coupled crystalline solids have shown that they can display various types of nontrivial 
bulk topological bands, \cite{KaneHasan_RMP2010,QiZhang_RMP2011,Konig02112007, Hsieh2008,weyl1,weyl2,weyl3,Chang2015, Bestwick2015} 
with unusual chiral surface states. Topologically nontrivial Chern bands have also been realized in recent experiments 
on cold atoms in two-dimensional (2D) lattices, using fermionic $^{40}$K atoms in a Floquet
 realization of the honeycomb lattice Haldane model with zero net flux per unit cell, \cite{ZhaiFloquet2014,Jotzu2014} and
 bosonic $^{87}$Rb atoms in a Hofstadter model with a net flux of $\pi/2$ per unit cell realized using Raman-assisted tunneling. \cite{Bloch15}
These developments have led to rejuvenated interest in understanding interaction effects in such topological bands. 

Early
work on the time-reversal symmetry broken Hofstadter model in an optical lattice showed that interacting bosons could realize
lattice analogues of various fractional quantum Hall (FQH) states, e.g., the $\nu\!=\!1/2$ Laughlin liquid, as well as FQH  states which have no continuum
analogues.\cite{Mueller2004,Sorensen2005, Palmer2006, Hafezi2007, Palmer2008, Moller2009,Mueller2010} More recently, attention has focussed on interaction effects in
Chern insulators. These break time-reversal symmetry and support bands with nonzero Chern numbers,
but possess full lattice translation symmetry due to {\it zero} net magnetic flux per elementary 
lattice unit cell (modulo an integer number of flux quanta per plaquette which can be gauged away).
In the regime of nearly flat Chern bands, where such bands mimic continuum Landau levels, interactions have been shown to drive
various types of FQH liquids for both spinless bosons and spinless fermions; \cite{Sheng2011,Regnault2011,fci_1,fci_2,fci_3,Wang11}
for recent reviews, see Refs.~\onlinecite{fci_rev1,fci_rev2}.
Similar issues have been previously studied also for interacting
time-reversal invariant topological insulators, such as the Kane-Mele-Hubbard model,\cite{Rachel2010,ZhengKMH_PRB2011,HohenadlerKMH_PRL2011,CenkeKMH_PRB2012,HsiangHsuanKMH_PRB2013,HohenadlerKMH_PRB2012}
as well as more realistic models appropriate to spin-orbit coupled oxides.\cite{FieteQSHI_PRL2012,LeHur_PRB2013} In these cases it has been
found that correlation effects lead to various easy-plane magnetic orders or $Z_2$ quantum spin liquids.
 
Here, we study the topological honeycomb lattice Haldane model \cite{Haldane1988} of a quantum anomalous Hall insulator (QAHI).
Previous work on spinless bosons in thie model at fractional filling found evidence for a fractional quantum Hall liquid. \cite{Wang11} At
 integer filling, it was found that weak interactions induce
 unusual superfluid phases, while strong interactions lead to plaquette Mott insulators with loop currents. \cite{Vasic15} On the other hand,
spinless fermions with nearest neighbor repulsion lead
to a topologically trivial charge-density wave crystal. \cite{Varney10,Varney11}
In recent work, it has been shown that the physics of two-component
 fermions (i.e., spin-$1/2$ fermions) in the Haldane-Hubbard model could be far richer. \cite{He2011_1,He2011_2,Maciejko13,Huber2014,Hickey2015,Zhai2015,HCPP2015} 
 Without interactions, this model has a total Chern number $C=2$
 at a filling of one fermion per site.
Within a strong coupling approach, the effective
 spin model of the Haldane-Mott insulator at this filling hosts chiral three-spin interactions.
 At the classical level, this model supports nontrivial chiral magnetic orders, including a triple-$Q$ tetrahedral phase; these orders
were shown to persist even for spin $S=1/2$ via an exact diagonalization (ED) study.\cite{HCPP2015} 
Furthermore, ED and density matrix renormalization
 group (DMRG) computations found that frustration-induced melting of the
 tetrahedral spin crystal resulted in a chiral spin liquid, \cite{Hickey2015,HCPP2015} the $\nu=1/2$ bosonic
 Laughlin liquid with topological degeneracy and gapped semion excitations, \cite{AndersonSL, KalmeyerLaughlin, Wen1990} 
This provides an example of a topological
 Mott insulator \cite{Pesin2010,Bhattacharjee12,Kargarian2012,Maciejko2014} in a realistic Hamiltonian.

 \begin{figure}[tb]
\includegraphics[width=\columnwidth,scale=0.25]{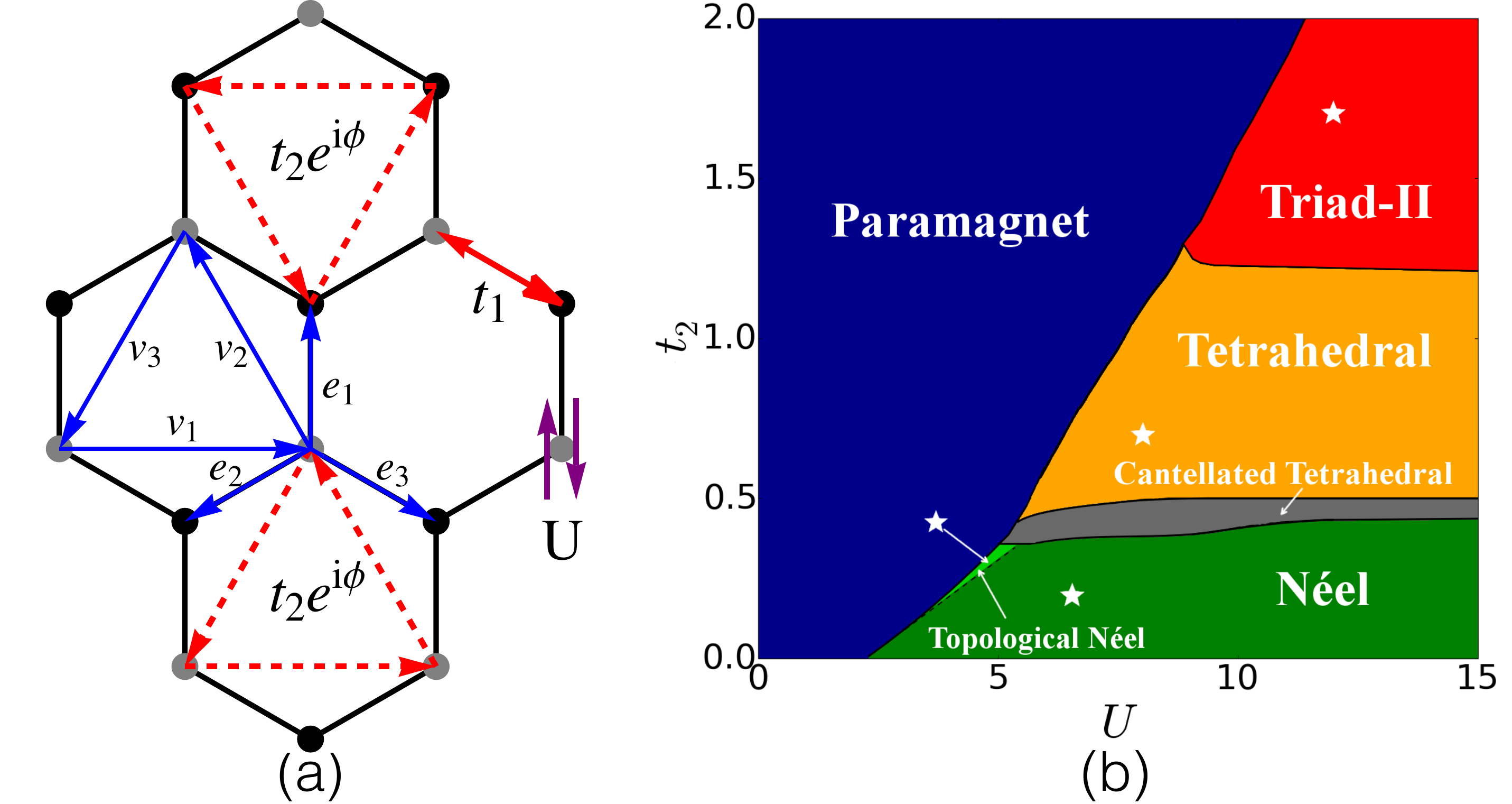}
\caption{(Color online) (a) Honeycomb lattice showing the two sublattices, the nearest and next-nearest neighbor lattice vectors $\{\vec{e}_i\}$ and $\{\vec{v}_i\}$, the hopping amplitudes $t_1, t_2$, phase $\phi$ and the Hubbard repulsion $U$ of the Haldane-Hubbard model. 
(b) Mean field phase diagram at $\phi=\pi/2$ showing the various magnetic phases that emerge at finite $U$. Stars ($\star$) mark the points at which the 
mean field band structure is plotted in Figs. \ref{fig:bands_Neel}, \ref{fig:bands_Tetra} and \ref{fig:bands_Triad}.}
\label{fig:lattice}
\end{figure}
 
 In this paper, we revisit the Haldane-Hubbard model away from its strong-coupling limit, and obtain the following key results.
 (i) Using an unrestricted real-space mean field theory of the Hubbard interaction, we examine all possible magnetic instabilities of
 the paramagnetic QAHI in an unbiased manner. Over a small window of parameters, we find the previously reported 
 topological N\'eel ordered phase, \cite{Zhai2015} which gives way to a topologically trivial N\'eel phase at larger interaction. More interestingly, 
 and reassuringly,
various chiral noncoplanar magnetic orders, such as the triple-$Q$ tetrahedral phase, previously
discovered within a strong coupling approach as arising from chiral three-spin interactions in the Mott insulating phase,
 also appear in this mean field theory. Our study provides estimates for the interaction strength at which 
 the QAHI becomes unstable to various broken symmetry insulating phases. 
While the emergence of noncoplanar chiral magnetic order at moderate coupling agrees with a previous mean-field study of the 
Haldane-Hubbard model,\cite{Zhai2015}
the ordering patterns we find in our work are considerably different from Ref.~\onlinecite{Zhai2015} since they
focused on a specific ansatz, whereas we do not make any {\it a priori} assumptions about the unit cell.
 (ii) Using the converged mean field solution, we 
 compute the band structure and Berry curvature of the various magnetically ordered insulating phases. 
 In the tetrahedral phase, we find that a na\"ive application of a commonly employed projection
approach, which assumes that the itinerant fermions in the ordered phase move adiabatically with spins
projected along the local Zeeman field axes, \cite{YBKBerry1999,TrivediProj2015} while it is generally a useful
approach, is {\it inadequate} in this case to understand the band dispersion since it 
produces spurious band touchings due to an artificial restoration of time-reversal symmetry. Our numerical
mean field calculations in the tetrahedral phase show that these band touchings in fact
get gapped out, and a Berry curvature computation yields robust topological
bands with high Chern numbers. 

Our work thus sheds light on the interaction-driven instabilities of quantum 
Hall band insulators towards Mott insulators with exotic magnetic orders, which may ultimately be driven into chiral spin liquids by 
frustration effects. In some ways, our study is a parallel of previous work on the Kane-Mele-Hubbard model \cite{Rachel2010}
where weak and strong coupling approaches were applied to understand the impact of interactions on a 2D  time-reversal invariant
topological insulator.

How relevant are these results to the ultracold atom experiments which realize the Haldane model as an effective description
of a Floquet problem associated with a driven optical lattice? Recent work on interaction effects in driven Hamiltonians have suggested that the Haldane-Hubbard
model and its strong coupling limit
may be a reasonable description of the physics in the regime where the drive frequency exceeds the Hubbard
model parameters, $\Omega \gg U,t$, and further that the heating rate is exponentially small
in this regime.\cite{Abanin2015a,Bukov2015,Abanin2015b} 
Indeed, the experimentally
measured heating rate for interacting two-component fermions appears not to be too large.\cite{Jotzu2014} 
In this case, equilibriation in the effective equilibrium description of the interacting Floquet problem could potentially occur (i.e., prethermalization)
before the onset of significant heating.\cite{Abanin2015a,Bukov2015,Abanin2015b} Thus, we expect atleast
short-range spin correlations of these magnetically ordered states to be visible using techniques such as Bragg scattering \cite{Corcovilos2010,Hulet2015}
and noise measurements.\cite{Altman2004}

\section{Model and mean field theory} 

The Haldane-Hubbard model for spin-$1/2$ fermions shown in Fig.~\ref{fig:lattice}(a) is defined by the Hamiltonian
\begin{align}
\notag H_{\rm HH} \! =\! & -t_1\! \!\! \sum_{\la i j \ra \sigma} (c^\dg_{i\sigma} c^\pdg_{j\sigma} \!+\! h.c.) 
\!-\!  t_2 \!\!\!\! \sum_{\la\la i j \ra\ra \sigma} \!\! (e^{i\nu_{ij}\phi} c^\dg_{i\sigma} c^\pdg _{j\sigma} \!+\! h.c.)  \\
&+ U \sum_{i} n_{i\uparrow}n_{i\downarrow},   \label{eqn:HaldaneHubbard}
\end{align}
where $\la.\ra$ and $\la\la.\ra\ra$ denote, respectively, first and second nearest neighbors, $\nu_{ij}=\pm 1$
produces a flux pattern with a net zero flux per unit cell, and $U$ is the Hubbard repulsion. 

Without interactions, $U\!=\! 0$, this model supports Chern bands for $t_2,\phi \neq 0$. At half-filling, this leads to a 
QAHI with $\sigma_{xy}=\pm e^2/h$ per spin
for small $|t_2|$. At large $|t_2|$ and $\phi\neq\pi/2$, the Chern bands strongly disperse, leading to a metal with 
$\sigma_{xy} \neq 0$ but non-quantized.\cite{Hickey2015} In this paper, we will focus attention on the case $\phi=\pi/2$, for
which the half-filled state is a QAHI for arbitrarily large $t_2$. What we would like to explore is the fate of this topological
band insulator as we crank up interactions.

For strong interactions, $U \gg t_1,t_2$, and a half-filled lattice with one fermion per site,  
it is obvious that the ground state is a charge-localized Mott insulator. The effective model
describing the residual spin degrees of freedom in the Mott insulator features antiferromagnetic 
Heisenberg interactions at ${\cal O}(t^2/U)$, and nonvanishing chiral three-spin
interactions at ${\cal O}(t^3/U^2)$, of the type $\vec S_1 \cdot \vec S_2 \times \vec S_3$ on triangular plaquettes.
The chiral terms originate from broken time-reversal symmetry. The competition between these two types
of interactions in the Mott phase has been shown to lead to a variety of chiral magnetic orders within a classical spin approximation,
as well as from ED studies on 32-site clusters. \cite{HCPP2015} However, this previous work does not shed light on the interaction strength at which these magnetic orders emerge, or whether truncating the large-$U$ expansion at ${\cal O}(t^3/U^2)$ is justified.

Here we therefore explore an alternative route to treating the Hubbard repulsion, using a traditional mean-field 
decoupling of the local four-fermion interaction into 
all possible channels, leading to an effective quadratic Hamiltonian. Previous studies along these lines have focussed on only 
certain types of magnetic orders; here, we carry out a completely unbiased self-consistent study by using a site-dependent mean-field theory on
$L\times L$ lattices (with $2 L^2$ sites due to having two sites per unit cell) with system sizes upto $L=12$, backed up by momentum space computations on large lattices. 
This allows us to recover certain previously identified phases such as 
a topological N\'eel phase. In addition, we find large regions of the phase diagram which support various chiral magnetic orders 
which coincide with those deduced from
the strong coupling approach in the Mott insulating phase.

To decompose the Hubbard interaction via a site-dependent Hartree factorization, we define the mean fields $\rho_i \!=\! \avg{n_i}$ and
$\vec{m}_i \!=\! \frac{1}{2} \avg{c_{i \alpha}^\dagger \vec{\sigma}^\pdg_{\alpha\beta} c^\pdg_{i\beta}}$ at each site $i$, corresponding, respectively, 
to the local charge density and magnetization. Using these, the
mean field Hamiltonian is
\begin{align}
\notag H_{\rm MF} \! =\! & -t_1\! \!\! \sum_{\la i j \ra \sigma} (c^\dg_{i\sigma} c^\pdg_{j\sigma} \!+\! h.c.) 
\!-\!  t_2 \!\!\!\! \sum_{\la\la i j \ra\ra \sigma} \!\! (e^{i\nu_{ij}\phi} c^\dg_{i\sigma} c^\pdg _{j\sigma} \!+\! h.c.)  \\
&+ \frac{U}{2} \sum_i \rho_i \sum_{\sigma} c^\dg_{i\sigma} c^\pdg_{i\sigma}  - 2 U \sum_i \vec{m}_i \cdot \vec{S}_i, \label{eqn:HMF}
\end{align}
where we have dropped constant terms. To solve the mean-field theory, we have to demand self-consistency of these mean
field parameters $\{\rho_i , \vec m_i\}$. Denoting the eigenfunctions of $H_{\rm MF}$ as $\Psi_n(i,\sigma)$, 
this condition reduces
to 
\bea
\rho_i &=& \sum_{n} \sum_\alpha \Psi_n^*(i,\alpha) \Psi_n^\pdg(i,\alpha) f(\epsilon_n) \label{mfeq1}  , \\
\vec m_i &=&  \frac{1}{2} \sum_{n}\sum_{\alpha\beta} \Psi_n^*(i,\alpha) \vec\sigma_{\alpha\beta} \Psi_n^\pdg(i,\beta) f(\epsilon_n) , \label{mfeq2} 
\eea
where $\epsilon_n$ is the energy of the $n^{\rm th}$ eigenstate, and $f(\epsilon_n)$ is the Fermi-Dirac function. Here, we restrict
ourselves to zero temperature. We consider a lattice with $N$ honeycomb lattice unit cells, which leads to $2N$ lattice sites and a total of $4N$
single particle states including spin. Thus, at zero temperature, the effect of the Fermi-Dirac function at half-filling 
amounts to summing $n$ over $2N$ lowest energy states. Furthermore we  set $t_1=1$ for the remainder of the paper. 

\begin{figure*}[tb]
\includegraphics[scale=0.5]{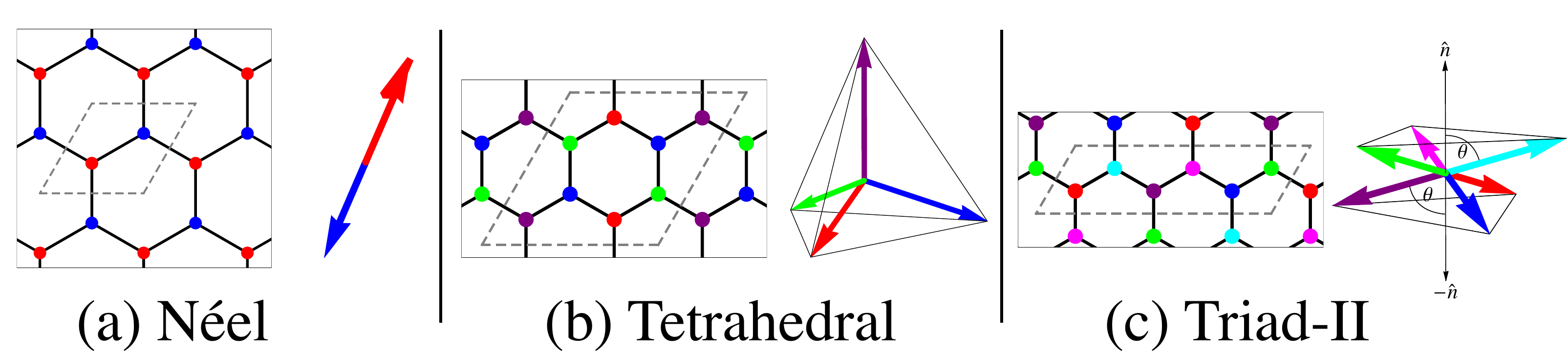}
\caption{(Color online) Real space spin structure of the (a) N\'eel, (b) tetrahedral and (c) triad-II phases. The sites are labelled by colors on the left side of each subfigure 
with the corresponding spin vector shown on the right side. The dashed lines mark out the magnetic unit cell.}
\label{fig:spins}
\end{figure*}

We have solved these mean field equations using an iterative self-consistent approach. 
Starting with an initial random guess for the local mean field parameters $\{\rho_i,\, \mathbf{m}_i\}$, 
we diagonalize $H_{\rm MF}$ in real space to find the eigenvalues and eigenvectors. We then use
these eigenvectors and energies to compute $\rho_i,\vec m_i$ via Eqns.~\ref{mfeq1},\ref{mfeq2},
iterating this process until convergence is reached. In our calculations we considered
$L\times L$ lattices (with $2 L^2$ sites) with system sizes up to $L=12$ with periodic 
boundary conditions. Since the local density rapidly converges
to a uniform value of one fermion per site in all cases, we found it suffices to impose the convergence 
condition on the magnetization $\sum_{i=1}^N |\Delta\mathbf{m}_i| < 10^{-6}$ where $\Delta\mathbf{m}_i$ 
is the change in $\mathbf{m}_i$ between subsequent iterations. 
As further checks on the phase diagram, we have used many different random initial conditions to start the self-consistency
loop. Furthermore, while $H_{\rm MF}$ in Eq. \ref{eqn:HMF} omitted constant terms which are unimportant for
self-consistency, we have re-instated this term to compute the total
mean field energy as
\begin{align}
\avg{H}= \sum_{n=1}^{2N} \epsilon_n + U \sum_{i=1}^{2N} \left(\mathbf{m}_i^2 - \rho^2_i/4\right).
\end{align}
We have explicitly checked that we are selecting ground states with the lowest energy $\la H \ra$.

\section{Magnetism and Band Topologies} 

Using the method described in the previous section, we were able to determine the phase diagram shown in Fig.~\ref{fig:lattice}(b). 
In addition to the paramagnetic QAHI, it features various magnetic phases which
we have labelled as topological N\'eel, N\'eel, tetrahedral, 
triad-II, and cantellated tetrahedral. Fig.~\ref{fig:spins} shows the magnetic configuration for the N\'eel, tetrahedral and triad-II phases.  All of the phases were found to have a
site-independent density $\rho_i =1$, so we focus here purely on their magnetic structure.  

Once we know the real space magnetic structure we can use the corresponding magnetic unit cell, see Fig. \ref{fig:spins}, to rewrite the Hamiltonian in momentum space. An example of this 
approach is given below for the case of N\'eel order. This approach also dramatically reduces the number of variational
degrees of freedom,
down to just one parameter in the case of the N\'eel and tetrahedral, and two parameters in the case of the triad-II. We use the same form of iterative self-consistent approach to 
determine the parameters as outlined in the previous section. This leads to excellent agreement between the real space and momentum space calculations. Computing the energy of each 
order directly in momentum space
on large lattices allows us to confirm the location of the various phase boundaries in Fig.~\ref{fig:lattice}(b). Furthermore, we use this to 
carry out Berry curvature and Chern number computations of the bands in the various phases, 
following the numerical procedure outlined by Fukui, Hatsugai and Suzuki. \cite{Fukui2005}

\subsection{Paramagnet}

\begin{figure}[tb]
\includegraphics[width=\columnwidth]{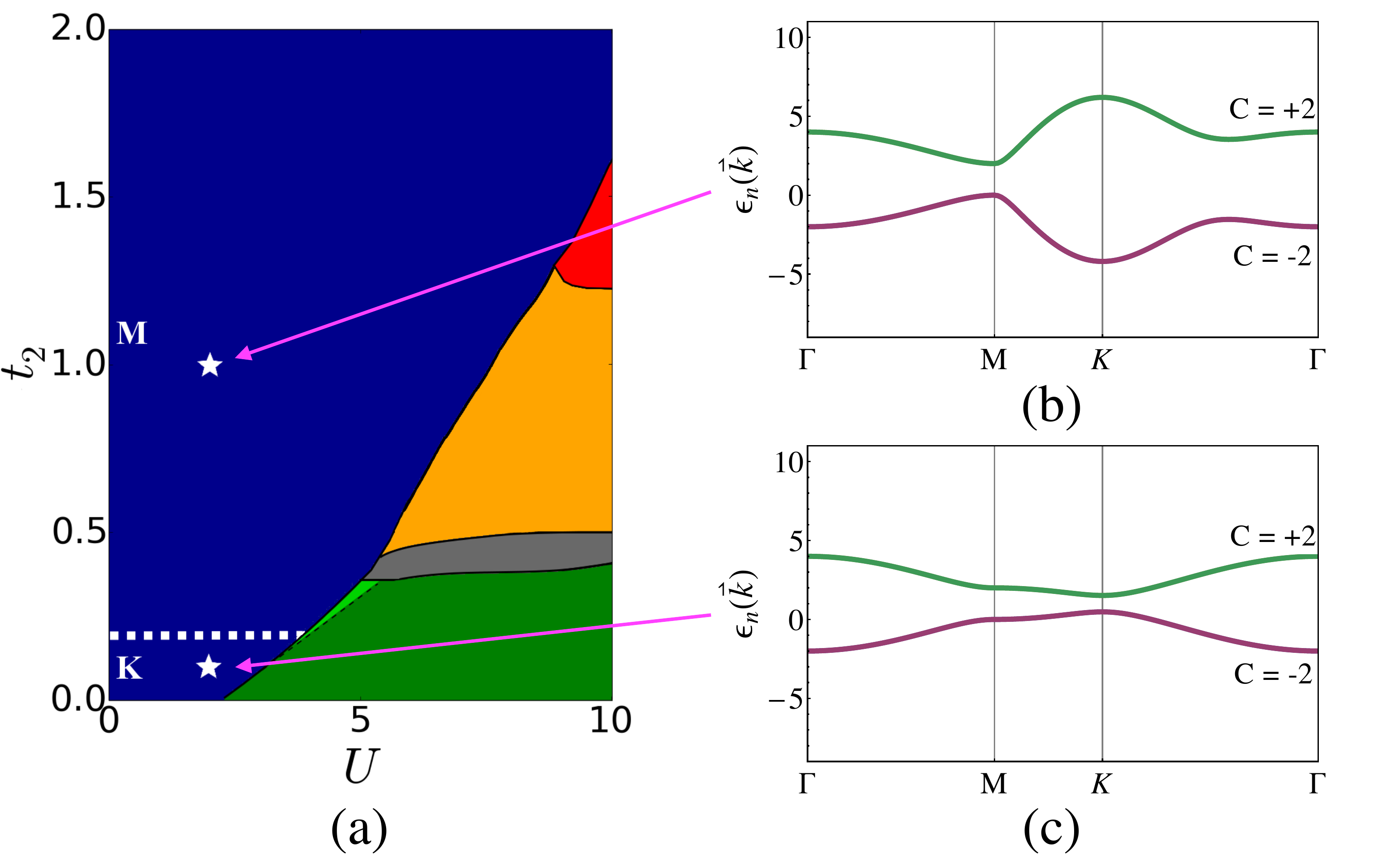}
\caption{(Color online) (a) Paramagnetic region of the phase diagram with the smallest band gap lying at the $M$ points above the dashed white line and at $\pm K$ below it. Examples of the 
band structure are shown at (b) $t_2=1.0$ and (c) $t_2 = 0.1$.}
\label{fig:bands_Para}
\end{figure}

The paramagnet corresponds to $\vec m_i \!=\! 0$ and $\rho_i\!=\! 1$. In mean field theory, this simply leads back to the 
non-interacting Haldane model. The inclusion of spin results in four bands, a set of two lower degenerate bands with 
a combined Chern number $C=-2$ and two
upper degenerate bands with $C=2$. For $t_2=0$ the upper and lower bands touch at Dirac points located at $\pm K$. 
A nonzero $t_2$ gaps out these band touchings, with the location of the minimum
band gap shifting from the $\pm K$ points to the $M$ points for $t_2 \!>\! 1/3\sqrt{3}$, as indicated by the dashed white line in Fig. \ref{fig:bands_Para}.

\subsection{N\'eel and Topological N\'eel}

The magnetic order in the N\'eel and Topological N\'eel states corresponds to collinear 
spins, aligned antiparallel on the two sublattices of the honeycomb lattice. In this state, $\la
\hat{\chi}_\triangle \ra=0$ on all triangular plaquettes, where $\hat{\chi}_\triangle \equiv \vec m_i \cdot \vec m_j \times \vec m_k$ is the scalar spin chirality operator, with the sites $\{ijk\}$ being
labeled going anticlockwise around the triangles. Its magnetic 
structure factor $ \vec{\mathcal{M}}(\bq)= \frac{1}{N} \sum_{i,j} \la \vec m_i \cdot \vec m_j \ra  {\rm e}^{i\bq\cdot(\vec r_i - \vec r_j)}$ exhibits a Bragg peak at $\bq=0$. 

To study the magnetic order in the N\'eel states in momentum space, we use the unit cell in Fig. \ref{fig:spins}(a) and orient the spins in the $\hat{\vec{z}}$ direction. 
The Fourier transformed N\'eel Hamiltonian has the form
\begin{align}
H_{\mathrm{N\acute{e}el}}=\sum_\vec{k} \vec{c}^\dagger_{\vec{k}} {\cal M}_{\mathrm{N\acute{e}el}}(\vec{k})\vec{c}^\pdg_{\vec{k}},
\end{align}
where $\vec{c}_{\vec{k}} = (c_{A,\uparrow}, c_{B,\uparrow}, c_{A,\downarrow}, c_{B,\downarrow})^\mathbf{T}$ with $A$,$B$ representing the two sublattices. 
In this basis, ${\cal M}$ is block diagonal,
\begin{align}
{\cal M}_{\mathrm{N\acute{e}el}}(\vec{k}) \!\!= \!\! \left(\begin{array}{cc}
\vec h(\vec{k})\cdot \vec{\tau} - Um\tau_z & 0 \\
0 & \vec h(\vec{k})\cdot \vec{\tau} + Um\tau_z \end{array}\right).
\label{MNeel}
\end{align}
Here, we have defined
\begin{align}
\!\! h_x(\vec{k}) \!\!&=\!\! -\left[\cos(\vec{k}\cdot\vec{e}_1) + \cos(\vec{k}\cdot\vec{e}_2) + \cos(\vec{k}\cdot\vec{e}_3)\right] ,\\
\!\! h_y(\vec{k}) \!\!&=\!\! -\left[\sin(\vec{k}\cdot\vec{e}_1) + \sin(\vec{k}\cdot\vec{e}_2) + \sin(\vec{k}\cdot\vec{e}_3) \right] ,\\
\!\! h_z(\vec{k}) \!\! &=\!\! -2t_2 \left[\sin(\vec{k}\!\cdot\!\vec{v}_1) \!+\! \sin(\vec{k}\!\cdot\!\vec{v}_2) \!+\! \sin(\vec{k}\!\cdot\!\vec{v}_3) \right],
\end{align}
with $\pm m \hat{z}$ being the magnetization on the $A$,$B$ sublattices, while the lattice unit vectors are given by
\bea
\!\! \vec{e}_1 \!\!&=&\!\! (0,a);~
\!\! \vec{e}_2 \!=\! (-\frac{\sqrt{3}a}{2},\,-\frac{a}{2});~
\!\! \vec{e}_3 \!=\! (\frac{\sqrt{3}a}{2},\,-\frac{a}{2}) ,\\
\!\! \vec{v}_1 \!\!&=&\!\! (\sqrt{3}a,0);~
\!\! \vec{v}_2 \!\!=\!\! (-\frac{\sqrt{3}a}{2},\,\frac{3a}{2});~
\!\! \vec{v}_3 \!\!=\!\! (-\frac{\sqrt{3}a}{2},\,-\frac{3a}{2}).
\eea

\begin{figure}[tb]
\includegraphics[width=\columnwidth]{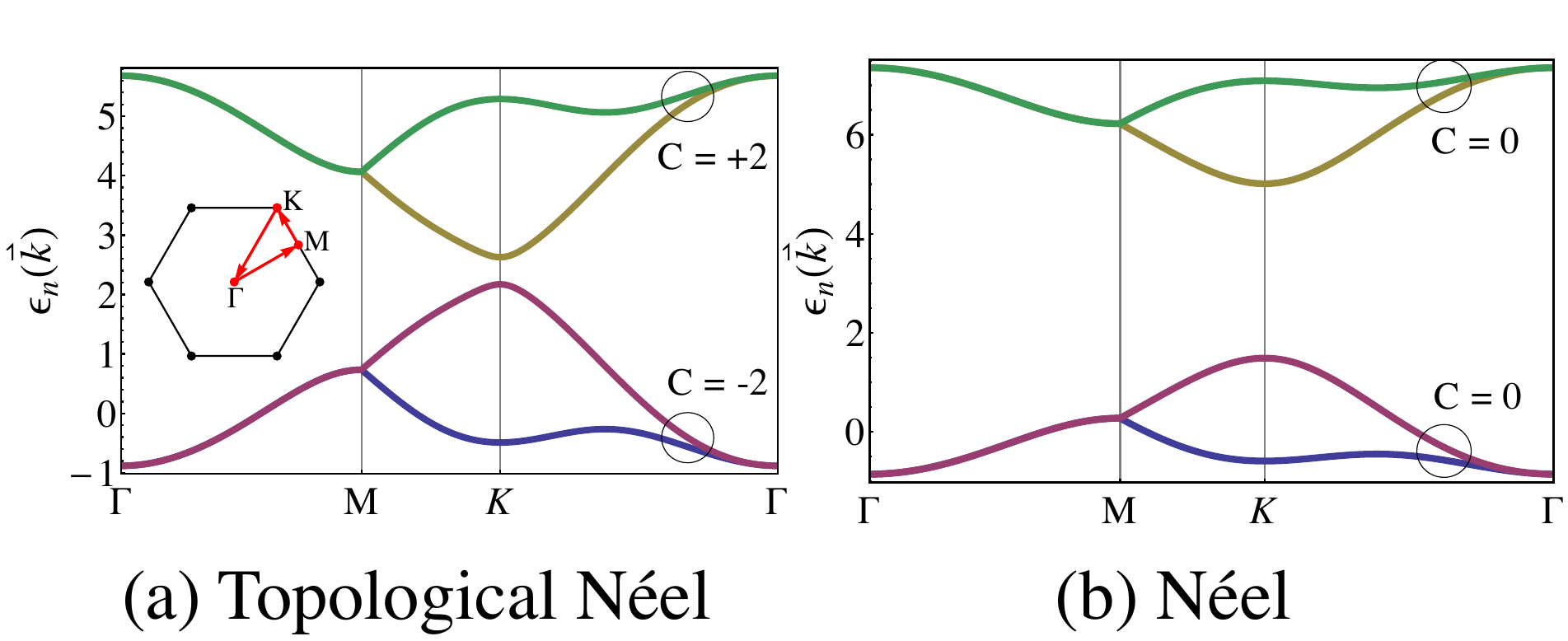}
\caption{(Color online) Band structure of the (a) topological N\'eel phase for $t_2=0.3$, $U=4.8$ and (b) N\'eel phase for $t_2=0.2$, $U=6.5$ along the high symmetry lines of 
the honeycomb lattice BZ (shown in the inset). The bands carry a total Chern number 
$C\!=\! \pm 2$ in the topological N\'eel phase, while $C\!=\!0$ in the N\'eel phase.}
\label{fig:bands_Neel}
\end{figure}

Solving the resulting self-consistent equation for $m$, we 
find that the onset of N\'eel order appears via a continuous transition from the paramagnet, with no closing of a single-particle gap. 
This transition may be viewed as a 
condensate of direct excitons formed from particles and holes at the $K$ point (or $-K$ point). Close to the transition, on the
ordered side,
the sublattice magnetization $\vec m$ is small, increasing continuously in magnitude
away from the critical point. We can find this critical repulsion for the onset of N\'eel order by expanding the
self-consistent equation to linear order in $m$, which yields 
\begin{align}
  \frac{1}{U_c} = \frac{1}{2L} \sum_{\substack{k \in \mathrm{BZ}}} \frac{(h_x^2 + h_y^2)}{|\vec{h}|^3}
\end{align}
where $L^2$ is the number of unit cells. When $m$ is small, as we can see from ${\cal M}_{\mathrm{N\acute{e}el}}(\vec{k})$ in Eq.~\ref{MNeel}, it leads to a small sublattice imbalance potential
for each spin;\cite{Zhai2015} this breaks the spin degeneracy present in the paramagnetic band structure. However, there exists a small window
of $U$ over which the topological properties of the paramagnetic bands survive, namely a combined $C\!=\! -2$ for the two lower bands and $C\!=\!+2$ for the two upper bands. Increasing $U$ drives a gap closing at the
$\pm K$ points which makes the combined Chern numbers of both sets of bands trivial. Thus, as noted in Ref. \onlinecite{Zhai2015},
 the band structure of the N\'eel allows us to divide it into topologically trivial and non-trivial
phases, with examples for each phase given in Fig. \ref{fig:bands_Neel}. The topological N\'eel 
to ordinary N\'eel transition is a topological transition with a change of band topology but no change in the sublattice magnetisation. However, as pointed out in Ref. \onlinecite{Zhai2015}, fluctuations beyond mean-field theory 
due to the coexistence of gapless fermions and gapless N\'eel spin wave modes may render the transition first-order. 

\subsection{Tetrahedral}

The tetrahedral has spins pointing from the origin towards the four corners of a tetrahedron, and
are tiled on the lattice as shown in Fig.~\ref{fig:spins}(b). This non-coplanar state has 
a uniform chirality $\la \hat{\chi}_\triangle \ra$ on each small-$\triangle$ formed by two nearest neighbor and 
one next-nearest neighbor link.
The tetrahedral state is a triple-$Q$ state, with the structure factor $\vec{\mathcal{M} }(\bq)$ exhibiting 
Bragg peaks at the $M$ points in the Brillouin zone (BZ). The transition into the tetrahedral state is first order, which we attribute to
symmetry-allowed third order chiral terms in a Landau theory formulation; we also confirm this with a 
direct numerical evaluation of the energy as a function of the magnetization.

\begin{figure}[tb]
\includegraphics[width=\columnwidth]{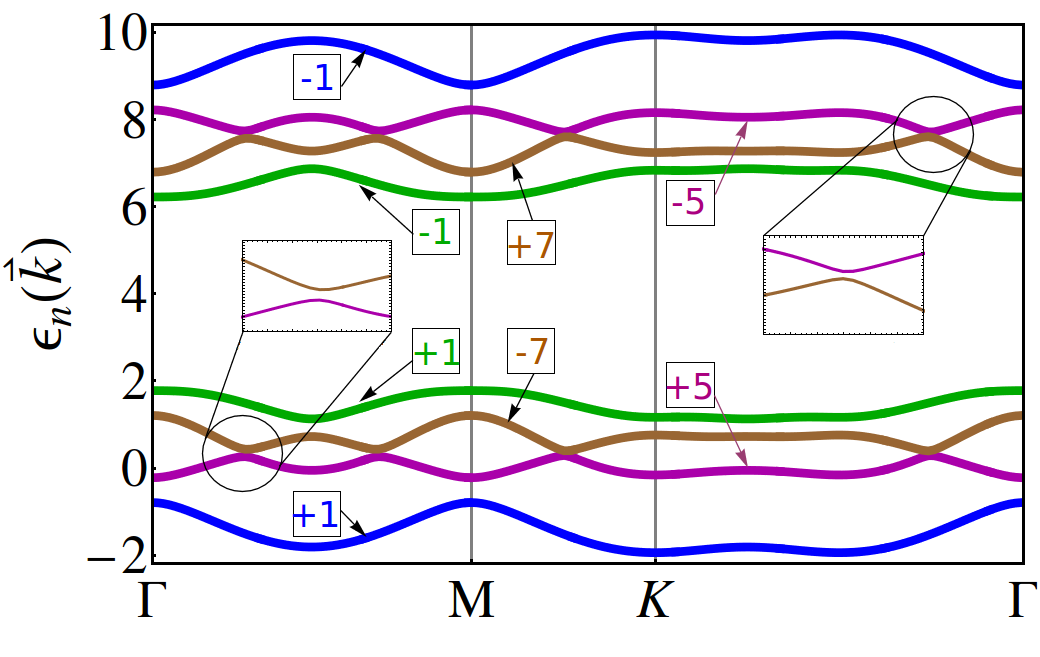}
\caption{(Color online) Band structure of the tetrahedral phase for $t_2=0.7$, $U=8$ along the high symmetry lines of 
the original honeycomb lattice BZ (shown in the inset of Fig.~\ref{fig:bands_Neel}(a)). Each band is doubly degenerate and there are no band touchings; since the gaps 
at a number of points are difficult to resolve by eye, we have magnified a couple of examples. Each set of degenerate bands is labelled by their total Chern number.}
\label{fig:bands_Tetra}
\end{figure}

The $8$-site magnetic unit cell of the tetrahedral results in a total of $16$ bands; these form $8$ degenerate pairs as seen from 
Fig.~\ref{fig:bands_Tetra} where we have plotted them in the original large BZ of the honeycomb lattice. The pairs of degenerate bands do not
intersect each other. This allows us to assign each pair a total Chern number as indicated in Fig.~\ref{fig:bands_Tetra}, computed
using the method outlined in Ref.~\onlinecite{Fukui2005}. With increasing $U$,
we find that the  gaps between these bands appears to close at a number of points in the BZ. However, the gaps always remain nonzero but
are difficult to resolve by eye, as we illustrate in the insets. The existence of these `almost band touching' points can be understood if we appeal to a `projection' approach
\cite{YBKBerry1999,TrivediProj2015}
which assumes that the bands in the magnetically ordered state can be constructed by studying fermions moving adiabatically in the background of the 
tetrahedral order, while having their spins perfectly polarized along the local Zeeman field axes.

In the projection approach, we ignore the explicit underlying tetrahedral order, instead assuming its effects are encoded in the emergent Berry phase factors \cite{YBKBerry1999,TrivediProj2015} for fermions
as they traverse closed loops
on the honeycomb lattice. Fermions hopping in a closed three-site loop will pick up a Berry phase of $\pm \pi/2$ corresponding to half the solid angle 
subtended by any three of the four spins of the tetrahedral. Combined with the Haldane flux $\phi = \pi/2$ this produces an effective flux of $\pm\pi$ through each small triangle as
well as a total flux of $\pm \pi$ through each hexagon. This modified flux pattern 
means that the projected system has an artificial time-reversal symmetry, that is not strictly present in the original Hamiltonian. Thus, the projection approximation
cannot lead to isolated gapped bands with
nonzero Chern numbers. There are only two other possibilities. (i) It could be that the band structure in this approximation has
isolated bands with $C=0$. However, since this adiabatic approximation is expected to
be good at large $U$, this is at odds with our mean field result. (ii) The bands in this approximation cannot be isolated and lead to band touchings. 
Indeed, we find that the presence of this artificial 
symmetry results in spurious symmetry-protected band touching points which coincide with the weakly gapped points in the full calculation shown in
Fig. \ref{fig:bands_Tetra}.
At any finite $U$, the `perfectly polarized' approximation is inadequate due to small but nonzero mixing with the higher energy states,
and this small deviation leads to tiny band gaps which we observe in the full mean field calculation in the
mean field tetrahedral phase; a computation of the Berry curvature  \cite{Fukui2005} in this mean field solution yields the observed Chern bands.
This illustrates that the projection approach, while it is widely employed and useful
for understanding gapped bands and their topological properties, should be used with care when it results in band touchings.

The tetrahedral state also features bands with high Chern numbers, namely $C=\pm5$ and $C=\mp7$. This is mainly due to a concentration of Berry curvature at six points in the 
BZ along the $\Gamma$-$K$ high symmetry line. We associate these large Berry curvature regions with
band touchings in the projection approach. Focusing on the middle two pairs of bands in the lower set each point contributes a Berry flux 
of $\pm 2\pi$, due to the twofold degeneracy, giving a total contribution of $C=\pm6$ to the total Chern number. For the lower pair this, added to a Berry flux contribution of $-2\pi$ 
from the $\Gamma$ point, results in $C=5$ while for the upper pair, added to a contribution of $+2\pi$ from $\Gamma$ and a total of $-4\pi$ from $\pm K$, results in $C=-7$.

\subsection{Triad-II}

The triad-II can be thought of as a cone, or umbrella, state on each triangular $A$, $B$ sublattice 
with the common axis of each cone, $\pm\hat{n}$ in Fig. \ref{fig:spins}(c), aligned anti-parallel and spins making an angle $\theta$ with $\pm\hat{n}$ (whereas the case in 
which the common axes align parallel is referred to as triad-I). It has a net anti-ferromagnetic moment 
and a structure factor $\vec{\mathcal{M} }(\bq)$ exhibiting Bragg peaks at $\Gamma$ and $\pm K$. Each cone can in general be rotated by an 
angle $\varphi$ about $\pm \hat{n}$ with respect to the other. The ground state energy has a weak sinusoidal dependence 
on $\varphi$ with minima at $\varphi\!=\!(2n+1)\pi/3$, and which goes to zero with increasing $U$. This differs from the results of the recent classical analysis of the strong-coupling spin model in which the energy of the 
triad-II is independent of $\varphi$. 

Fixing $\varphi\!=\!\pi/3$ in the momentum space calculations leaves two independent variational parameters, the magnetization per site $m$ and the opening angle of the cones, $\theta$. For fixed $t_2$ the opening angle is an increasing function of $U$, approaching the limiting value of $\pi/2$ in which the 
spins on each triangular sublattice form an incommensurate coplanar spiral state. At fixed $U$ the variation of $\theta$ with $t_2$ is relatively negligible for the range of $t_2$ studied here. We note that, on large 
system sizes, the triad-II state is expected to be unstable to very weak incommensurate spiralling. \cite{Hickey2015} 

The transition from the paramagnet into the triad-II state is first order, which we again attribute to third order chiral terms in a Landau theory formulation. We have again confirmed with a 
direct numerical evaluation of the energy.

\begin{figure}[tb]
\includegraphics[width=\columnwidth]{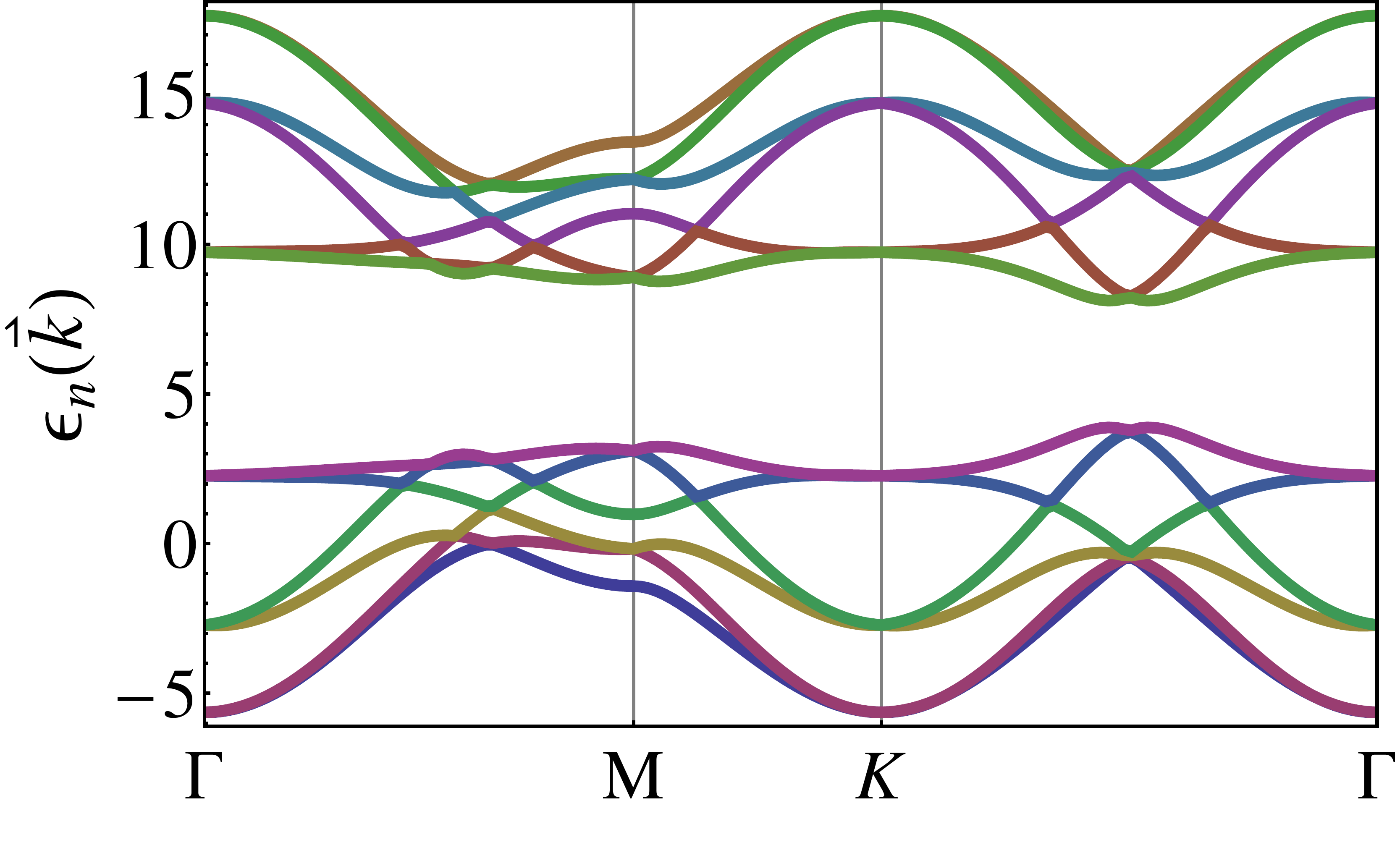}
\caption{(Color online) Band structure of the triad-II phase for $t_2=1.7$, $U=12$ along the high symmetry lines of 
the original honeycomb lattice BZ (shown in the inset of Fig.~\ref{fig:bands_Neel}(a)). None of the bands in the lower or upper set are separated in 
energy and thus it is not meaningful to label the bands with Chern numbers. The total Chern number of the lower (upper) six bands is zero.}
\label{fig:bands_Triad}
\end{figure}

The $6$-site magnetic unit cell of the triad-II results in a total of $12$ bands, as shown in Fig. \ref{fig:bands_Triad} in the
original large BZ of the honeycomb lattice There are no bands separated in energy in the upper/lower set so it is not possible to 
assign Chern numbers to any one band. The total Chern number of the lower six bands (or upper six bands) is zero.

\subsection{Cantellated Tetrahedral}

\begin{figure}[tb]
\includegraphics[width=\columnwidth]{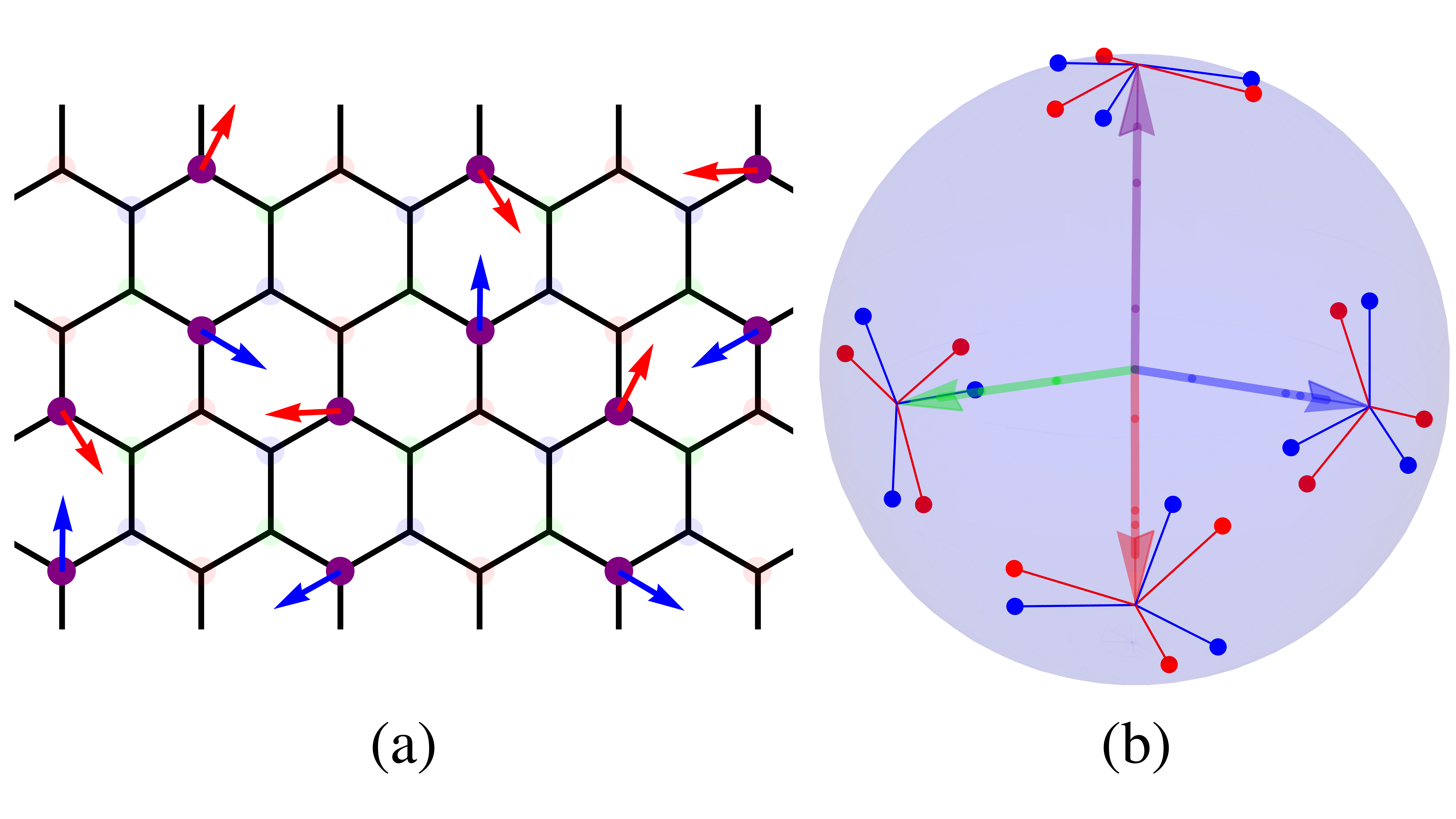}
\caption{(Color online) In the tetrahedral state, the spins pointing towards the four vertices of the tetrahedron
each reside on their own honeycomb lattice in real space, in (a) the purple dots highlight the sites corresponding to one of the vertices. 
In the cantellated tetrahedral each tetrahedral vertex splits into six points, with the spins residing on one (e.g. purple) sublattice developing
nonzero transverse (tangential)
components, shown by the red/blue arrows in (a). These transverse components form $120$ degree ordered states on the two triangular 
sublattices of the purple honeycomb lattice. In (b) the 
points correspond to the vertices of the cantellated tetrahedral order, with the blue/red color distinguishing the two sublattices of the honeycomb 
lattice and the arrows indicating parent tetrahedral vertices.}
\label{fig:Cantellated}
\end{figure}

The Cantellated Tetrahedral (CT) state descends from a parent tetrahedral state. In the parent tetrahedral configuration, the spins marked with a given color
in Fig.~\ref{fig:spins} form a honeycomb lattice, with a 
larger unit cell. We can deform these ferromagnetically aligned spins by dividing the larger honeycomb lattice into 
two sublattices, and then allowing for a $\sqrt{3}\times\sqrt{3}$ canting on each sublattice; this splits each
vertex of the tetrahedron into six vertices (forming a small hexagon) leading to the CT state with a $24$-site unit cell; see
Fig. \ref{fig:Cantellated}. 

The CT state exists over a narrow window between the N\'eel and Tetrahedral states, indicated by the gray region of Fig. \ref{fig:lattice}.
The real space self-consistent calculations did not converge to the required accuracy at all the points in the region 
marked as Cantellated Tetrahedral. However on repeating the calculations using this phase as the initial condition, the algorithm converged 
at many more points within the gray region to the CT phase. Even at points within the gray region where the solution did not 
converge to the desired accuracy, the resulting magnetic configuration closely resembled that of the CT phase even though it was not precisely the same, 
and the energies of these states were lower than that of the other phases. We thus tentatively associate the CT phase with the entire gray region.
Due to uncertainties associated with this state, the small window over which it might appear, 
and the much larger size of its unit cell, we have not further explored its bands and their topological character.

\section{Comparison with previous work}

We have explored the possible density and magnetic instabilities of the half-filled Haldane-Hubbard model. Using large, unbiased site-dependent 
mean field theory calculations in real space lattices, we have uncovered a variety of magnetic orders, N\'eel, cantellated tetrahedral, tetrahedral and triad-II. 
The non-coplanar orders found in our study differ from \onlinecite{Zhai2015}, since that work only explored a limited ansatz, whereas we do not make
any such assumptions. Our ordered phases are in good agreement with 
those found in recent classical and ED studies of the effective spin model that emerges from a strong coupling approach. \cite{Hickey2015,HCPP2015}
(although the cantellated tetrahedral could not be identified in the ED study due to the size of the finite clusters used).
As $U$ increases even further, the chiral terms that appear in the spin model at ${\cal O}(t^3/U^2)$ become much weaker, and the effective spin
model tends towards the $J_1$-$J_2$ model on the honeycomb lattice. Classically this model has N\'eel and coplanar spiral ground states which 
again match with our mean-field results on the Hubbard model at much larger values of $U$.

Although the mean-field approach captures the magnetic orders seen in the strong coupling limit it cannot be relied upon to completely capture the
physics near the Mott transition into magnetically ordered insulating states. Spin and charge fluctuations beyond mean field theory may 
modify the nature of these transitions, as pointed 
out in Ref. \onlinecite{Zhai2015} for the case of the topological N\'eel to N\'eel transition - this transition is continuous in our mean field calculations
but is predicted to be driven first-order by fluctuation effects. The transitions from the paramagnetic QAHI into the
tetrahedral and triad states are found to be first-order in our calculations, driven by cubic chiral terms; this result is unlikely to be qualitatively
affected by
the inclusion of beyond-mean-field fluctuations.

\section{Summary}

In summary, we have explored the unconventional magnetic structures found in the Haldane-Hubbard model, and 
explored the band structure of the resulting phases. This allows us to identify when the paramagnetic quantum Hall insulator
becomes unstable to magnetically ordered insulators, to uncover the topologically non-trivial region of the 
N\'eel phase as well as the high Chern numbers of the tetrahedral phase, and to make connections with previous studies of this model
in the strong coupling limit.

{\it Note Added:} After submission of this work for publication, some 
recent works have appeared  Refs.~\onlinecite{MaciejkoDec2015,LiDec2015,TroyerDec2015}  which study the Haldane-Hubbard
model using different approximations such as dynamical mean field theory and the variational cluster approximation. In particular,
Ref.~\onlinecite{TroyerDec2015} studies the effect of sublattice potential imbalance which 
we have not considered in our work, and finds the emergence of an interaction-induced $C=1$ Chern insulator.

\noindent{\bf Acknowledgements.} We thank L. Cincio, R. Desbuquois, G. Jotzu, and Z. Papic for useful 
discussions. We acknowledge support from NSERC of Canada.

%

\end{document}